# A Survey of Green Networking Research

Aruna Prem Bianzino, Claude Chaudet, Dario Rossi, Jean-Louis Rougier
Institut TELECOM, TELECOM ParisTech, CNRS LTCI UMR 5141, Paris, France
Email: {bianzino, chaudet, drossi, rougier}@telecom-paristech.fr

*Abstract*—Reduction of unnecessary energy consumption is becoming a major concern in wired networking, because of the potential economical benefits and of its expected environmental impact. These issues, usually referred to as "green networking", relate to embedding energy-awareness in the design, in the devices and in the protocols of networks.

In this work, we first formulate a more precise definition of the "green" attribute. We furthermore identify a few paradigms that are the key enablers of energy-aware networking research. We then overview the current state of the art and provide a taxonomy of the relevant work, with a special focus on wired networking. At a high level, we identify four branches of green networking research that stem from different observations on the root causes of energy waste, namely (i) Adaptive Link Rate, (ii) Interface proxying, (iii) Energy-aware infrastructures and (iv) Energy-aware applications. In this work, we do not only explore specific proposals pertaining to each of the above branches, but also offer a perspective for research.

*Index Terms*—Green Networking; Wired Networks; Adaptive Link Rate; Interface Proxying; Energy-aware Infrastructures; Energy-aware Applications

## I. INTRODUCTION

The reduction of energy consumption has become a key issue for industries, because of economical, environmental and marketing reasons. If this concern has a strong influence on electronics designers, the information and communication technology sector, and more specifically the networking field, is also concerned. For instance, data-centers and networking infrastructure involve high-performance and high-availability machines. They therefore rely on powerful devices, which require energy-consuming air conditioning to sustain their operation, and which are organized in a redundant architecture. As these architectures are often designed to endure peak load and degraded conditions, they are under-utilized in normal operation, leaving a large room for energy savings. In recent years, valuable efforts have indeed been dedicated to reducing unnecessary energy expenditure, which is usually nicknamed as a *greening* of the networking technologies and protocols.

As energy-related studies in wireless networks are very specific and would require a dedicated study, this survey focuses on *wired* networks, even though a section exposes some wireless technologies optimizations. In these networks, energy saving often requires to reduce network performance or redundancy. Considering this compromise between the network performance and energy savings, determining efficient strategies to limit the network energy consumption is a real challenge. However, although the green networking field is still in its infancy, a number of interesting works have already been carried out, which are overviewed in the present survey. The rest of the paper is organized as follows. Section II motivates the case for energy-awareness in the telecom infrastructures, concluding with a formal problem statement. Section III overviews the general paradigms that can be exploited to reduce energy consumption. Section IV introduces the criteria for a taxonomy of the green networking solutions, which we then apply to survey the state of the art in Sections V–VII. Section IX instead briefly overviews the current research trends in areas that are closely related to, but out of the scope of this survey – namely, *computer* and *data-center* architectures and *wireless* networking. The problem of network energy measurement is addressed in Section X, devoting special care to issues tied to the comparison of different solutions. Finally, Section XI presents a summary along with some interesting open directions, which current work have only partly dealt with, and which we believe future work should address.

## II. MOTIVATIONS AND OBJECTIVES

### A. Why save energy

Consciousness of environmental problems tied to Green-House Gases (GHG) increased during the recent years. All around the world, various studies started highlighting the devastating effects of massive GHG emissions and their consequences on the climate change. According to a report published by the European Union [1], a decrease in emission volume of 15%–30% is required before year 2020 to keep the global temperature increase below 2°C.

GHG effects are not limited to the environment, though. Their influence on economy have also been investigated and their financial damage has been put in perspective with the potential economical benefits that would follow GHG reduction. In particular, [2] projected that a 1/3 reduction of the GHG emissions may generate an economical benefit higher than the investment required to reach this goal. Political powers are also seeking to build a momentum around a greener industry, both in the perspective of enforcing a sustainable long-term development, and as a possible economic upturn factor on a shorter perspective.

GHG reduction objectives involve many industry branches, including the Information and Communication Technology (ICT) sector, especially considering the penetration of these technologies in everyday life. Indeed, the volume of $CO_2$ emissions produced by the ICT sector alone has been estimated to an approximate 2% of the total man-made emissions in [3]. This figure is similar to the one exhibited by the global airline industry, but with higher increase perspectives. Moreover, when considering only developed countries such as the United Kingdom, this figure rises up to 10% [4].

As the precise evaluation of these numbers is a difficult process, these projections are likely neither entirely accurate, nor up-to-date. Nevertheless, these studies all agree on the fact that ICT represents an important source of energy consumption and GHG emissions. Even if the incentives are still not clear (e.g., in term of regulations), there seems to be a clear innovation opportunity in making network devices and protocols aware of the energy they consume, so that they can make efficient and responsible (or "green") decisions.

*B. Where to save energy*

Before attempting to reduce energy consumption, or to understand by what means such reduction can be achieved, it is necessary to identify where the largest improvements could take place. The Internet, for instance, can be segmented into a core network and several types of access networks. In these different segments, the equipment involved, its objectives and its expected performance and energy consumption levels differ. As such, one may reasonably expect that both the consumption figures and the possible enhancements are considerably different. In 2002, [5] analyzed the energy consumption contributions of different categories of equipment in the global Internet. These figures, represented in Fig. 1, indicate that local area networks, through hubs and switches, are responsible for about 80% of the total Internet consumption at that time. In 2005, the authors of [6] estimate the relative contribution of the Network Interface Cards (NICs) and all the other network elements and conclude that the NICs are responsible for almost half of the total power consumption. More recently, studies have started suggesting an increase of the consumption in the network core: for instance, in 2009 Deutsche Telekom [7] forecasts that by year 2017, the power consumption of the network core will be equal to that of the network access (the study also suggested a stunning 300% rise in power consumption of the network core in coming decade).

Not surprisingly, everything evolves rapidly in the ICT domain, which makes the aforementioned figures and estimations quickly out-dated and possibly inaccurate. As a consequence, there is a true need for a permanent evaluation of this consumption, in order to point out and update regularly the most relevant targets for potential energy-savings. However, such an evaluation requires a collaboration of equipment manufacturers, Internet service providers and governments, which is clearly not an easy process. We will come back on this issue in Section XI.

*C. Definition of green networking*

From a strict **environmental point of view**, the objective of green networking is to aim at the minimization of the GHG emissions. An obvious first step in this direction is to enforce as much as possible the use of renewable energy in ICT. Yet another natural track is to design low power components, able to offer the same level of performance.

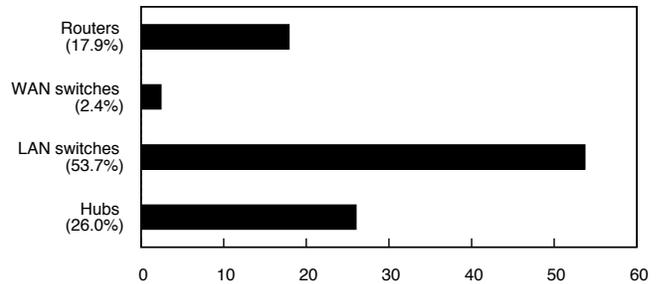

Fig. 1. Contribution of different device types to the network energy consumption in year 2002 [5]

However, this is not the only lead: redesigning the network architecture itself, for instance by de-locating network equipment towards strategic places, may yield substantial savings too for two main reasons. The first reason is related to the losses that appear when energy is transported: the closer the consumption points are to the production points, the lower this loss will be[1]. The second reason is related to the cooling of electronic devices: air cooling represents an important share of the energy expenditure in data centers and cold climates may loosen this dependency. Canada advanced research and innovation network (CANARIE)[8] is strongly pushing in this direction, especially using virtualization to ease service geographical delocalization driven by the energy source availability.

Furthermore, large ICT companies like Google displaced their server farms to the banks of the Columbia River to take advantage of the energy offered by the hydroelectric power plants nearby. The water flow provided by the river may in addition be used within the cooling systems, as experimented by Google [9], even though this may lead to other environmental issues such as seaweed proliferation if the water temperature increases too much[2]. An alternative cooling system, investigated by Microsoft in the *In Tent* and *Marlow* projects [10] consists in leaving servers in the open air so that heating dissipates more easily.

Geographical delocalization is also a promising approach from an **economical point of view**. The global energy market offers volatile and time-varying prices. The prices may even become negative when a production surplus appears but there is no customer demand. Energy cannot be stored efficiently, and even though consumption predictions based on historical data quite accurately trigger production units, over-production is always possible. This variability can be exploited by displacing the computation where energy has a lower cost. Companies like Amazon explore such geographical delocalization of services [11] in order to reduce the operational expenditures related to energy supply.

---

[1]Long-distance electricity transportation uses high-voltage lines to reduce losses: therefore, the energy losses are not directly proportional to the distance, but may rather be represented by a threshold-based linear function.

[2]Such considerations lead to a strict surveillance of nuclear power plants for instance.

Going one step further, if the physical machines can be delocalized to minimize the global energy consumption, one may imagine that services too may be located at the optimal places, and that they may move when conditions vary, introducing the time dimension. CPU-intensive operations may be executed on one hemisphere or on the other so that CPU cycles follow, e.g., seasonal or day/night patterns. A huge technological challenge lies in performing such service migration without any service disruption, preserving fault-tolerance and data security.

The previous optimizations are directly function of the energy price and are not directly related to environmental considerations. The market-related issues behind this problem may lead to an optimal solution in terms of cost that is sub-optimal in terms of total energy consumption. Indeed, 100 MWh sold at a unit price of $120 are more expensive than 120 MWh sold at a unit price of $90. Thus, environmental considerations generally need a **regulator point of view** to assist their enforcement. Regulation, which often falls into governmental duties, may push towards greening of the technology by different means (e.g., taxes on GHG emissions, diverting research funds towards energy efficiency, etc.).

Finally, from an **engineering point of view**, green networking may be better seen as a way to reduce energy required to carry out a given task while maintaining the same level of performance, which is the point of view that we will adopt in the rest of this article. Nevertheless, this point of view alone is still relevant as system efficiency from the engineering perspective still deeply relates to economical, regulatory and environmental viewpoints.

## III. GREEN STRATEGIES

Traditionally, networking systems are designed and dimensioned according to principles that are inherently in opposition with green networking objectives: namely, *over-provisioning* and *redundancy*. On the one hand, due to the lack of QoS support from the Internet architecture, over-provisioning is a common practice: networks are dimensioned to sustain peak hour traffic, with extra capacity to allow for unexpected events. As a result, during low traffic periods, over-provisioned networks are also over-energy-consuming. Moreover, for resiliency and fault-tolerance, networks are also designed in a redundant manner. Devices are added to the infrastructure with the sole purpose of taking over the duty when another device fails, which further adds to the overall energy consumption. These objectives, radically opposed to the environmental ones, make green networking an interesting, and technically challenging, research field. A major shift is indeed needed in networking research and development to introduce energy-awareness in the network design, without compromising either the quality of service or the network reliability.

This section illustrates a few key paradigms that the network infrastructure can exploit to reach the green objectives formalized above. We individuate four classes of solution, namely **resource consolidation**, **virtualization**, **selective connectedness**, and **proportional computing**. These four categories represent four research directions, which may find further detailed applications in device and protocol design.

**Resource consolidation** regroups all the dimensioning strategies to reduce the global consumption due to devices underutilized at a given time. Given that the traffic level in a given network approximately follows a well-known daily and weekly behavior [11], there is an opportunity to "adapt" the level of active over-provisioning to the current network conditions. In other words, the required level of performance will still be guaranteed, but using an amount of resources that is dimensioned for current network traffic demand rather than for the peak demand. This can, for example, be achieved by shutting down some lightly loaded routers and rerouting the traffic on a smaller number of active network equipment. Resource consolidation is already a popular approach in other fields, in particular data centers and CPU.

Applying the same base concept, **selective connectedness** of devices, as outlined in [12], [13], consists in distributed mechanisms allowing single pieces of equipment to go idle for some time, as transparently as possible for the rest of the networked devices. If the consolidation principle applies to resources that are shared *within* the network infrastructure, selective connectedness allows instead to turn off unused resources *at the edge* of the network. For instance, edge nodes can go idle in order to avoid supporting network connectivity tasks (e.g., periodically sending heartbeats, receiving unnecessary broadcast traffic, etc.). These tasks may have to be taken over by other nodes, such as proxies, momentarily faking identity of idle devices, so that no fundamental change is required in network protocols.

**Virtualization** regroups a set of mechanisms allowing more than one service to operate on the same piece of hardware, thus improving the hardware utilization. It results in a lowered energy consumption, as long as a single machine under high load consumes less than several lightly loaded ones, which is generally the case. Virtualization can be applied to multiple kinds of resources, including network links, storage devices, software resources, etc. A typical example of virtualization consists in sharing servers in data centers, thus reducing hardware costs, improving energy management and reducing energy and cooling costs, ultimately reducing data center carbon footprint. In this context, virtualization has already been deployed with success: e.g., the US Postal Service has virtualized 791 of its 895 physical servers [14]. As virtualization is a more mature research field, we refer the interested reader to [15] for a detailed survey of virtualization techniques from a computer architecture perspective, and to [16] for a networking perspective. At the same time, it should be noted that a virtualization solution designed explicitly to reduce network energy consumption has yet to appear.

**Proportional computing** was introduced in [17] and may be applied to a system as a whole, to network protocols, as well as to individual devices and components. To illustrate this principle, Fig. 2(a) depicts different energy consumption (or cost) profiles that a device may exhibit as a function of its utilization level (or demand). Utilization and energy

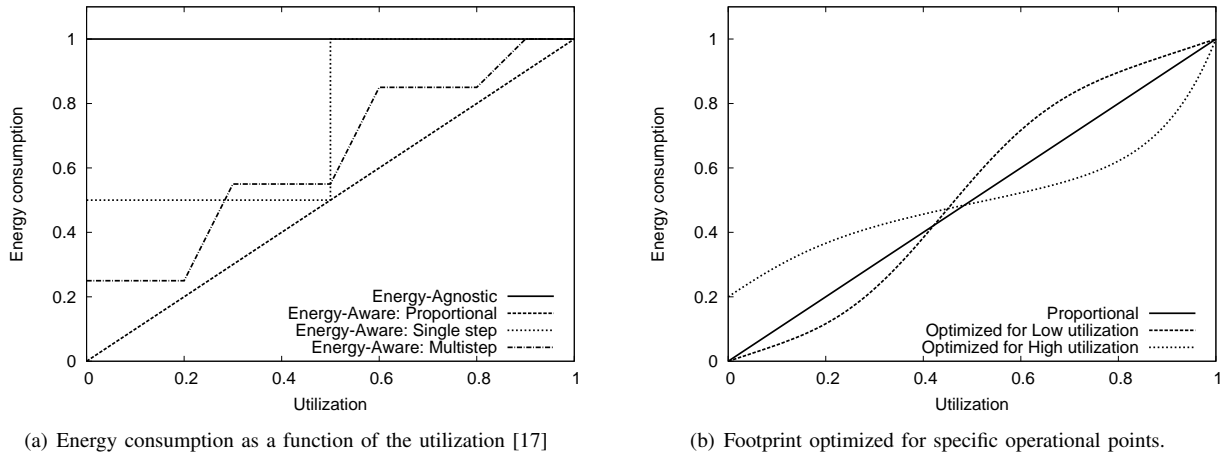

(a) Energy consumption as a function of the utilization [17]

(b) Footprint optimized for specific operational points.

Fig. 2. Representation of Green footprint of a networked system, protocol or device.

metrics have been normalized and the axis range from 0 to 1. These different profiles offer different optimization opportunities. *Energy-agnostic* devices, whose energy consumption is constant, independently of their utilization, represent the worst case: such devices are either on and consume the maximum amount of energy, or off and inoperative. In contrast, fully *energy-aware* devices exhibit an energy consumption proportional to their utilization level. Between these two extreme situations, there exist an infinite number of possible intermediate profiles, for instance the *single-step* and *multi-step* cases represented in Fig. 2(a), whose energy consumption coarsely adapts to their load. Single step devices have two operation modes while multi-step devices have several performance thresholds. Dynamic Voltage Scaling and Adaptive Link Rate are typical examples of proportional computing. Dynamic Voltage Scaling [18] reduces the energy state of the CPU as a function of a system load, while Adaptive Link Rate (that we extensively analyze in Section V) applies a similar concept to network interfaces, reducing their capacity, and thus their consumption, as a function of the link load.

It should be noted that an additional challenge lays in the selection and enforcement of the target loads over a distributed architecture, as the overall energy consumption depends on the utilization level at which every energy-aware device operates. For instance, Fig. 2(b), represents two different footprints besides the proportional one. One of them is more apt for devices/protocol operating normally under higher utilization levels, and the other is more efficient for lower operational points. This latter design pattern can be motivated by the fact that, rather typically, both data-centers and network backbones are estimated to be most of the time running at low utilization levels. However, following the resource consolidation principle, it is likely that future systems receive incentives to operate at higher loads. Such interaction adds a further interesting dimension that may have to be considered: we shall not only design adaptive systems based on different levels of utilization, but we may also design systems optimized for the *most likely* utilization level.

## IV. Taxonomy of Green Networking Research

While, for the time being, network devices and protocol are mostly unaware of the energy they consume, a number of valuable research works have started exploring energy-awareness in fixed networks. In this section, we introduce different general criteria, which are summarized in Table I and that can assist us in classifying the different green techniques proposed so far in the literature.

### A. Classification criteria

A first important criterion concerns the **timescale** of the decisions involved by the green strategy. As pointed out in [12], timescales on the order of nanoseconds to microseconds apply to CPU and instruction level, which is relevant in the computer and software architecture levels, and thus only concern individual building blocks of a single system. Timescales on the order of micro to milliseconds are instead relevant at the system layer. At these timescales, actions may be taken between consecutive packets of the same flow (inter-packets, intra-flow), possibly involving several components at the same time, but likely confined within a single system. Larger timescales, on the order of one second and above, allow instead the action to span between multiple entities, possibly involving coordination of such entities as well. Notice that timescales directly define the architectural level at which actions can be taken: the shorter the timescale is, the lower the layer will be and the less possible interaction among different components will exist.

To simplify the taxonomy, in the following we adopt a coarse differentiation, only separating *on-line* solutions from *off-line* ones. The first set refers to the solutions thought to act at run time (with timescales shorter than one second). The second set refers instead to solutions acting before the run time (with timescales above the second), as for instance during the network design process. More precisely, depending on the type and amount of entities involved, we further divide *on-line* solutions as either *local* or *global*, depending on the scope of the information required to take a decision. Local strategies will require information that pertains to a single node, or single link, while global strategies will require information that pertains to a set of nodes and links. We also further differentiate between *off-line* solutions pertaining to *network design* (e.g., related to the choice of resources that will be employed in the network) and to *network dimensioning* (e.g., which consist in deciding how to use these resources, once they have been chosen in the previous step).

TABLE I
SUMMARY OF THE CLASSIFICATION CRITERIA

| Criterion | Values | Comments |
|---|---|---|
| Timescale | Online vs. Offline | Defines the update frequency of the policy |
| Scope (online) | Local vs. global | Influences the volume of communication required to reach the objective |
| OSI Layer | Link, Network, Transport, Application, Cross | Individuate which entities shall collaborate |
| Input process | Instantaneous vs. Historical | Defines learning and adaptation capabilities of the algorithm |
| Approach | Traffic analysis, Modeling, Simulation, Hw/Sw Prototype | Flavor of the study, also correlates with the level of maturity of the work |

TABLE II
TAXONOMY OF GREEN NETWORKING RESEARCH. ITEMS WITH ★ APPEAR IN MULTIPLE CATEGORIES.

| Branch | Reference | Timescale | | Network Layer | Approach | Input Process | Comments |
|---|---|---|---|---|---|---|---|
| | | On-line | Off-line | | | | |
| Adaptive Link Rate (ALR) | ★[19] Gupta and Singh | Global | | Data Link | Traffic Analysis | Forecast | Position paper, motivates ALR, Coordinate Sleeping mode, based on Arrival Process (AP) |
| | [20] Gupta et al. | Local | | Data Link | Trace Driven Sim. | Forecast | Sleeping mode, AP |
| | ★[21] Gunaratne et al. | Local | | Data Link | Trace Driven Sim. | Instant. | ALR part of the paper; Rate Switching, based on the Queue occupancy (Q) |
| | [22] Gunaratne et al. | Local | | Data Link | Trace Driven Sim. | Instant. | Rate switching, Q |
| | [23] Gupta and Singh | Local | | Data Link | Traffic Trace | Forecast | Sleeping mode, Q and AP |
| | ★[24] Nedevschi et al. | Local | | Data Link | Simulation | Forecast | Sleeping mode vs Rate switching, Q and AP; Abilene topology |
| | [25] Ananthanarayanan and Katz | Local | | Data Link | Trace Driven Sim. | History | Switching architecture; Sleeping mode, AP |
| | [26] Gunaratne et al. | Local | | Data Link | Math Model & Sim. | History | Trace and synthetic traffic; Rate switching, Q |
| Interface Proxying | ★[21] Gunaratne et al. | Local | | Cross | Traffic Analysis | Instant. | Proxying part of the paper; motivates NIC and External |
| | [27] Purushothaman et al. | Local | | Application | Trace Driven Sim. | Instant. | NIC Proxying for Gnutella protocol |
| | [28] Jimeno and Christensen | Local | | Application | Hw prototype | Instant. | External Proxying for Gnutella protocol |
| | [29] Sabhanatarajan and Gordon-Ross | Local | | Network | Trace Driven Sim. | Instant. | NIC Proxying; packet inspection |
| | [30] Agarwal et al. | Local | | Cross | Hw prototype | Instant. | NIC Proxying; *Somniloquy* prototype |
| | [31] Nedevschi et al. | Local | | Cross | Sw prototype | Instant. | Trace Driven evaluation; External Proxying |
| Energy Aware Infrastructure | ★[19] Gupta and Singh | | Dim. | Network | Traffic Analysis | History | Position paper, motivates Energy Aware Routing |
| | ★[24] Nedevschi et al. | Global | | Data Link | Simulation | Instant. | Coordinate sleeping; Abilene topology |
| | [32] Baldi and Ofek | | Design | Cross | Architectural | | |
| | [33] Chiaraviglio et al. | | Dim. | Network | Numerical solution | History | Energy-aware routing; Heuristic solution of ILP problem |
| | [34] Chabarek et al. | | Design | Network | Operational Research | | Energy-aware infrastructure, Router power-profiling |
| | [35] Sansò and Mellah | | Design | Network | Operational Research | | |
| | [36] Da Costa et al. | Global | | Application | Trace Driven Sim. | Instant. | Grid5000 management |
| Energy Aware Applications | [37] Irish and Christensen | Local | | Transport | Sw Prototype | Instant. | TCP split connection; protocol modification |
| | [38] Wang and Singh | | Design | Transport | Sw Prototype | | TCP optimization of FreeBSDv5 |
| | [39] Blackburn and Christensen | Local | | Application | Sw Prototype | Instant. | Telnet protocol modification |
| | [40] Blackburn and Christensen | Local | | Application | Simulation (ns2) | Instant. | BitTorrent protocol modification |

As IP networks are implemented following a layering principle, we may also classify solutions according to the **networking layer** to which they apply. Considering the TCP/IP protocols stack, each solution can either be implemented at a single layer among *data-link*, *network*, *transport* and *application* layers, or may require *cross-layer* interaction.

Another classification criterion comes from the analysis of the **input process** that drives the decision taken in the solutions. The decision may be taken on the basis of the *instantaneous* situation, on the basis of an *historical* observation, or on the basis of a *forecast* (depending on both the instantaneous situation and on historical observations). In the case of online solutions, all the three kinds of decision are possible. For the off-line solutions, decisions based on the instantaneous situation clearly do not make sense.

Finally, another important criterion reflecting the level of realism of the proposed solutions is related to the methodology or **approach** employed to evaluate the proposal. By approach, we mean for example discrete event simulation, hardware prototyping or formal models with analytical or numerical solution. All methodologies have their respective merit, advantages and limits. For instance, theoretical analysis and simulation studies provide implementation-independent results, compared to prototype, hardware or real deployment-based approaches. Yet, we point out that the latter approaches may reflect a higher level of maturity of a specific research field.

Table I summarizes the criteria we chose to derive our classification. Other criteria intuitively come into mind, but are not very relevant in the context of this paper. For instance, a natural classification criterion could refer to the **network segment** where each technique may be applied, i.e., either access, metropolitan or core networks. However, as until very recently the largest gain was expected to be achieved at the network edge (i.e., access networks), most pieces of work we overview actually fit in this segment (unless explicitly stated). Another possible criterion concerns the **type of service** to which the technique applies. However, as green techniques are typically applicable to a rather general extent, in the following we prefer to adopt a service agnostic viewpoint. Nevertheless, we also consider works targeting very specific service when needed, as in sections VI and VIII.

*B. Classification proposal*

Table II presents, at a glance, the main works that we overview in the following sections. Each row in the table refers to a different contribution while columns contain the different criteria outlined above. Strategies are clustered together in different groups, or branches, identifying the main energy-aware strategies currently defined.

In Section V we first analyze solutions that are referred to as **Adaptive Link Rate**, to which most of the effort in green networking has been devoted up to now. These techniques, following the proportional computing paradigm, are designed to reduce energy consumption in response to low utilization in an on-line manner. Techniques can be either considered to be link-local or network-global depending on the network layer they pertain to, as well as on the scale of the network involved and the need for interaction between elements (in which case, they also apply the selective connectedness principle). A considerable number of works have explored this solution, and the IEEE Energy Efficient Ethernet Task Force is moving toward its standardization as IEEE 802.3az [41].

Then, in Section VI we consider a set of work inspired mainly by the selective connectedness principle, which we refer to as **Interface Proxying** and that seek a reduction of unnecessary energy waste of edge devices. The main idea in this case is to delegate network-related traffic processing from power-hungry mainboard CPUs to low-power devices onboard of Network Interface Cards (NIC) or to external proxy devices. Also this second branch has already been explored by a relatively large number of works, and is the object of standardization from the ECMA association [42].

We identify two further areas of research, which are inspired by several of the principles earlier outlined, and which remain much less explored. More precisely, Section VII focuses on **Energy-Aware Infrastructure**, while we highlight some emerging efforts concerning **Energy-Aware Applications** in Section VIII.

For completeness, we provide in Section IX a few references to areas that are closely related to, but not directly in the scope of this survey, such as data-center, computer architectures and wireless networking. Finally, we report in Section X on the measurement and modeling effort of energy consumption in networked system and equipment, which has lately received significant attention as well.

V. ADAPTIVE LINK RATE

Empirical measurement showed that energy consumption on an Ethernet link is largely independent of its utilization [34], [43], [44]. In practice, even during the idle intervals where no frame is transmitted, the links are used to continuously send meaningless traffic in order to preserve synchronization and avoid the time required to send a long frame preamble. Therefore, the energy consumption of a link largely depends on the negotiated link capacity rather than on the actual link load.

A number of works have thus proposed to adapt the link rate by either (i) turning off links during (possibly short) idle periods, which is usually referred to as *sleeping mode* or by (ii) reducing the line rate during low utilization period, which is known as *rate switching*. Fig. 3 illustrates these two strategies. In both cases, the link rate is adapted to match the real network usage, hence approaching an energy consumption proportional to the link utilization by an on-line local adaptation of link configuration. Referring again to Fig. 2(a), this solution aims at bringing the energy consumption of a link from the initial constant worst case, down to a curve closer to the ideal proportional case.

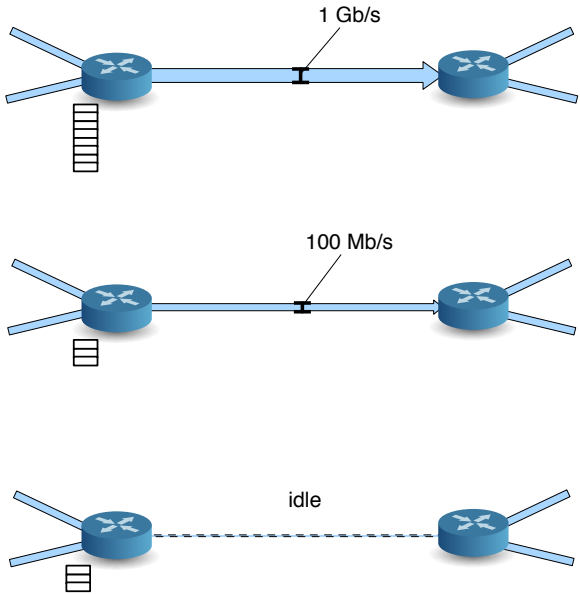

Fig. 3. Adaptive Link Rate strategies: the rate of a 1 Gb/s link can be reduced to 100 Mb/s (rate switch, middle plot) or the link can be made idle to save energy (sleeping mode, bottom plot), depending on the adjacent routers loads.

### A. Sleeping mode

A first subclass of adaptive link rate strategies includes simple energy-aware systems, such as [19], [20], [23], that only consider two states of operation: a sleeping (or idle) mode, and a fully working mode. The difficulty in this case consists in finding the desired compromise between system re-activity and energy savings. A recent survey covering sleeping mode from an algorithmic point of view can be found in [45] (which consider other computer science problems, such as job scheduling, that are however less close to networking).

In the pioneering work [19], the authors let the nodes decide on their interfaces status by measuring packet inter-arrival times, and considering if this interval is long enough to justify an effective energy saving between two consecutive frames. As the efficiency of such a strategy is directly tied to the inter-arrival distributions, the authors analyze a traffic trace to determine whether such an approach would be effective or not in practice.

Aside from the raw results, this first work raises a number of questions, addressed by subsequent research. First of all, different types of sleeping mode are possible for an interface, depending on the technology. The interface may (i) be in a deep idle state and drop the packets arriving during the sleeping period [23], or (ii) be fully awakened by every packet reception, or (iii) use a buffer to store the packets received during the sleeping intervals, processing them when it wakes up, or (iv) use a shadow port that may handle the packets on behalf of a cluster of sleeping ports [25]. However, even semi-sleep state does have a price in terms of energy. First, any idle state in which packets can be detected requires some electronics to be active, and thus consumes a small amount of energy, as well as powering any shadow port does. Hence, waking up the sleeping interface at every packet arrival reduces the packet delay and loss, but it will also reduce the energy savings.

In [20], the authors propose a two-state model of the sleeping mode strategy. The first state corresponds to the regular operation mode, and the second one to the energy saving mode. The first transition, from the energy saving mode to the operational mode, takes time (with the state of the art technology, the "wake-up time" is approximately 0.1 ms) and generates an energy consumption spike. The second transition, from the regular working mode to the energy saving mode, is supposed to be instantaneous and spike-less. This results in a simple, but not simplistic, model that can easily be extended to include more than two states, modeling, for instance, rate switch strategies.

### B. Rate switch

Besides the choice between an idle and a working mode, most of today's technologies propose a wider range of possibilities through the use of several transmission rates [21], [22], [26], to which different energy consumption figures correspond.

Ethernet, for instance, defines several transmission rates, from 10 Mb/s to 10 Gb/s, and higher capacities are under way. The authors of [21] show that there is a non-negligible difference in the interface energy consumption, across the different data rates. For instance, an increase of the data rate of a PC end system Network Interface Card (NIC) from 10 Mb/s to 1 Gb/s results in an increased energy consumption of about 3 W, which represents about 5% of the overall system energy consumption. For regular switches, the same throughput shift results in a per-interface energy consumption increase of about 1.5 W per link.

The same authors of [21] proposed successive refinement of the rate control policies, based only on the current system state [22], or on an historical analysis [26].

From a high-level point of view, selecting the proper data rate among a limited set of possibilities can be translated in an integer linear program whose objective is to minimize the overall energy consumption given that all the data inputted at the network is forwarded, which is known to be NP-hard.

### C. Comparison of sleeping mode and rate switch

In [24], the authors propose a comparison between a sleeping-mode and a rate switch algorithm, both applied at a network infrastructure level. The two algorithms are compared in terms of QoS, through the end-to-end packet delay and loss rate, and in terms of achieved energy saving, represented by the percentage of time in which network elements may sleep, in the *sleeping mode* case, or by the average rate reduction, in the *rate switch* case. From the energy saving point of view, the comparison highlights a network utilization threshold below which the *sleeping mode* performs better than *rate switch* and vice-versa. Moreover, the authors compare two types of rate sets. In the first case, the rates are distributed exponentially

(e.g., 10 Mb/s, 100 Mb/s, 1000 Mb/s) while in the second case they are distributed uniformly (e.g., 330 Mb/s, 660 Mb/s, 1000 Mb/s). Notice that the authors of that paper do consider linear rates (although they are not implemented in commercial IEEE 802.3 Ethernet cards), precisely to assess whether there could be further advantages in case hypothetical cards existed offering a linear scaling of the rates (as is roughly the case for IEEE 802.11 WiFi, which could motivate the manufacturer to produce such cards for Ethernet as well). Interestingly, results show that uniformly distributed rates perform better than the exponentially distributed ones, in terms of added delay and average rate reduction (hence, achieved energy saving). An increase in the number of supported rates is also shown to result in better performance, at the price of an increased management complexity.

Other works such as [46], [47] provide a relative comparison of sleeping mode versus rate switch strategies, when applied to processors and servers, respectively. In this case, both works are in favor of sleeping mode strategies because of a lower management complexity for a comparable performance level. The lower complexity comes with simpler optimization goals, which are minimizing idle energy and transition time, instead of complex load-proportional operation from each system component. Yet, dynamic rate switching is more robust in presence of bursty traffic, e.g., in case of mis-estimation of workload parameters. Interestingly, the above works reach the same conclusions from two rather different standpoints, since [46] adopts a prototyping approach, whereas [47] adopts a more theoretical one.

### D. Practical considerations

As we have seen, a subtle trade-off between energy savings and quality of service exists: moreover, network elements can trigger the adaptation of transmission rate based on different parameters.

The simplest way is probably to base the decisions on observations of the buffer state (Q in Table II), such as instantaneous measures of the buffer size. This solution has been examined by [21], which proposes to define two different buffer thresholds to drive the idle and active state changes. The use of two thresholds avoids too frequent oscillations between the two modes, but does not fully prevent such phenomena. In [26], the authors show how this solution may still provoke oscillations when the link utilization is close to the lower transmission rate. The authors propose a solution consisting in associating variable timers to dynamically adapt the thresholds. Timers are used to measure the amount of time spent in a given state: whenever state switching becomes too frequent, the corresponding threshold value is changed (e.g., doubled) to avoid rate oscillations. The state change may also be driven by an observation of the buffer state over a time window [25], or by a forecast of the future buffer states, on the basis of the present one and of the parameters of the arrival process (AP in Table II) [23]. These solutions lead to better performance, at the price of an increased system complexity.

Another aspect related to practical implementation issues concerns the synchronization of link terminations: i.e., once an interface decides to change its state, a mechanism is needed to make this change effective on both sides of the link. The authors of [19] propose a solution for the *sleeping mode*, in which an interface informs its neighbor just before entering sleeping mode, or sends a "wake up" packet to a sleeping neighbor to which it needs to send a packet. For the *rate switch*, [22], [26] point out how the rate Auto-Negotiation mechanism of Ethernet is unsuitable to data rate switching, because of its long action time (e.g., about 256 ms at 1 Gb/s), and propose a faster handshake based on MAC frames (i.e., request of rate change by one of the end systems, ACK/NACK reply by the other, taking less than $100\,\mu s$ at 1 Gb/s for handshake and resynchronization).

Finally, other practical aspects are considered by the IEEE Energy Efficient Ethernet Task Force, which is moving toward a standardization of the sleeping mode solution, including the definition of a low energy state for the intervals in which the link is underutilized [41].

## VI. INTERFACE PROXYING

The previous section introduced a set of mechanisms dedicated to put *interfaces* in idle mode: clearly, the longer the idle period is, the higher the achievable gain can be. It is thus worth assessing whether a similar approach could be applied to *end devices*, such as PCs.

With respect to the previous case, where functionalities could be simply turned off (e.g., no transmission at all when the link is idle), in the case of end devices this is however not possible, as some functionalities need to be delegated (i.e., traffic processing is handed over more energy efficient entities). Indeed, even though *users* are idle, background network traffic is nevertheless received and needs processing, preventing thus PCs from going in sleeping mode. The authors of [21], [31] point out that most of the incoming traffic received by a PC network interface during otherwise idle periods can simply be dropped or does not require more than a minimal computation and response. For instance, most broadcast frames or traffic related to port scanning may simply be ignored. Usual exchanges, such as ARP processing, ICMP echo answering or DHCP rebinding, are simple tasks that could be easily performed directly by the network interface. As in [21], we call *chatter* the received traffic that can be dropped without any noticeable effect.

The idea behind Interface Proxying consists in delegating the processing of such traffic. Processing can imply *plain filtering* (e.g., in the case of unwanted broadcast/portscan traffic) or may require *simple responses* (e.g., in the case of ARP, ICMP, DHCP), or even more *complex task* (e.g., in the case of P2P applications such as Gnutella or BitTorrent).

Such tasks can be delegated from the energy-hungry mainboard CPUs of end devices to a number of different entities: e.g., either instance locally to the low-energy processor onboard of the NIC of the same device, or to an external entity (in this latter case, a single proxy may be deployed for several machines in a LAN environment, or this duty may fall to the set-top-box in a residential environment). With respect to our taxonomy, proxying follows the principle of selective connectedness, and falls into the set of on-line local solutions.

## A. NIC proxying

NIC proxying implement a filtering or light processing of the received packets: the NIC may drop the chatter and handle the traffic requiring minimal computation, while the full system will be woken up only when non-trivial packets needing further processing are received. This allows energy saving through powering down the end systems, without disrupting their network connectivity. According to [21], this solution may apply to more than 90% of the received packets on a PC during idle periods: the authors propose longer sleeping intervals and a reduced number of system wake-ups for higher energy saving.

Another study on the achievable results and effects of proxying has been published in [27]. This work focuses on a solution that allows the computer system to enter a standby state, without losing its network connectivity. However, an analysis of the side effects of the time needed for the whole system to wake-up is missing, even though it is clearly not negligible,.

In [29], the authors analyze the possibility of implementing interface proxying, including traffic classification, over currently available hardware. The proposed framework supports a line speed of up to 1 Gbps in its software implementation (on the *Smart-NIC*), and up to 10 Gbps in its hardware implementation. The hardware implementation consumes only 25% of the energy required by the software-based one.

Extending this technique, [30] offloads some more complex tasks usually performed by the main CPU to the NIC. The authors show that it is possible, through the use of dedicated memory or direct memory access (DMA), to handle most of the network tasks that do not need interaction from the user, like Peer-to-Peer (P2P) applications, FTP downloads, status message update on Instant Messaging (IM) systems, etc. For instance, [30] augments NIC with an always-on low consumption CPU equipped with a small amount of RAM (64 MB) and flash storage (2 GB). Stub code, specific for each application that has to be offloaded from the main processor has then to be written. The authors implement the device and stub for Instant Messaging, BitTorrent and Web download.

## B. External proxying

Offloading traffic filtering and processing to an external machine may have several advantages in case of a larger LAN: besides the economy of scale, as the proxy acts for a number of end-devices, it can feature a more efficient CPU and thus offload the end-host from an even higher number of network-maintenance tasks. Fig. 4 illustrates such a situation in which a switch answers to an ARP request in place of the targeted computer, allowing this machine to remain asleep until it receives applicative traffic. For instance, beside ARP, ICMP, DCHP responses, a proxy can also maintain, for example TCP connectivity for the idle hosts: a splitting of TCP connections at the energy-aware proxy is for instance proposed in [21].

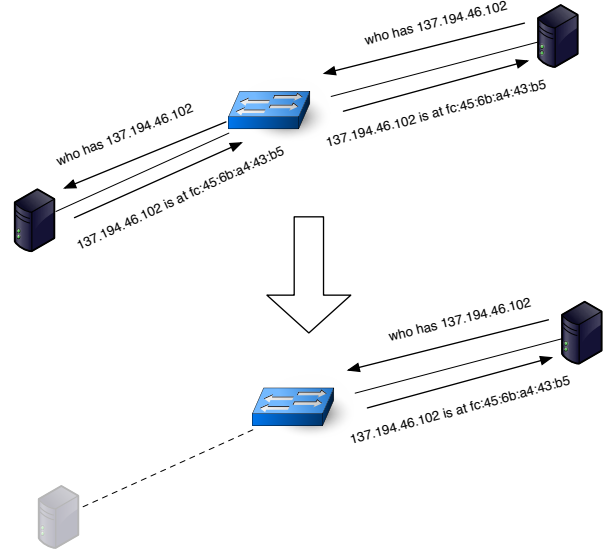

Fig. 4. External proxying: a switch acts as a proxy for ARP traffic, allowing the target machine to sleep at least until data traffic is sent.

External energy-aware proxies have also been evaluated in the context of P2P file-sharing applications [27], [28]. In P2P, edge device network presence represents a key issue to guarantee the robustness of the network: in this case, interface proxying represents a good way to save energy, without perturbing the system. The idea is explored in [27], with a prototype implementation for the Gnutella network in [28].

Energy-aware proxies are instead implemented in [31] as Click modular routers: the authors implement four different kinds of proxies, of increasing complexity, showing that, although the potential energy saving is considerable, nevertheless trivial approaches are not sufficient to fully exploit the potential saving. Indeed, while broadcast traffic is easily filtered, a significant implementation effort is needed to properly handle *unicast* traffic.

Finally, we note that all the above work do not take into account the residential environment, where set-top-boxes are likely to offer opportunities for external proxying.

## VII. ENERGY-AWARE INFRASTRUCTURE

The mechanisms seen so far only involve local decisions, through a single device or a very small set of collaborative devices. While these techniques alone offer non negligible energy saving, further improvement can be expected from a reasonable amount of collaboration between individual devices, sharing a wider knowledge on the system state.

## A. Energy-aware architecture

Two opposite approaches are possible to build an energy-aware architecture: an incremental approach, building over existing infrastructures, and a clean-slate approach, which advocates the complete redesign of a new architecture. So far, few works exist that have tackled energy-awareness from a global architectural perspective: [36], [24] prefer an incremental approach, while [32], [34], [35] follow a clean-slate design.

More precisely, [36] considers the problem of managing Grid5000 resources. Grid5000 is a grid platform used by researchers: prior to each experiment, a request for resources is issued, so that resources are attributed only to a single experiment at any time. However, as experiments may not need all the available resources, there may be resources that thus sit idle but remain powered on anyway. The authors propose a centralized global management approach, in which they schedule resource on/off periods according to the received request, and evaluate the energy savings by applying their scheme on a historical request record spanning over two years.

Another incremental approach is proposed in [24], which considers the automatic adaptation of the link rate from the global viewpoint of a whole backbone network. In detail, traffic is reshaped into bursts at the network edge, by arranging all packets destined to the same egress router to be contiguous within the bursts (similarly to what is done by optical burst switching techniques [48], [49]). This approach adds end-to-end delay, but only at the network ingress. Consequently, periods of activity and sleep alternate inside the network, allowing fewer state transitions with respect to a simple opportunistic "wake on arrival" strategy. The work studies the incidence on the time spent in sleeping mode for different parameters: average network utilization, burst size, and state transition time. The authors argue that this strategy does not add significant complexity to the network, although they do not propose strategies to determine when and for how long nodes should ideally sleep (notice that global coordination may be very hard to achieve). Moreover, the effects of the traffic shaping on the jitter are not analyzed, even if they are expected to be important.

Other works instead advocate the use of clean-slate approaches, with a higher use of optical networks (e.g. Dense Wavelength Division Multiplexing). It is now admitted that optical switching is much more energy efficient, while offering an extremely large capacity. At the same time, these technologies still suffer from a lack of flexibility with respect to the electronics domain (as in the optical domain no buffering is possible, which motivated optical burst switching [48], [49]). A future challenge is probably to find efficient architectures combining both optical transport and packet processing, when needed. For instance, [32] falls in this category by proposing to complement the Internet with a parallel virtual DWDM "super-highway" dedicated for deterministic traffic.

Finally, the problem of introducing energy-awareness into the network design process is studied in [34], [35] from an operational research point of view. In more detail, [34] introduces the energy consumption cost into the multicommodity formulation of the design problem, together with the performance and robustness constraints. A similar approach is adopted in [35], which evaluates also the tradeoff between energy consumption and network performance, highlighting the fault tolerance characteristics of the different possible working points.

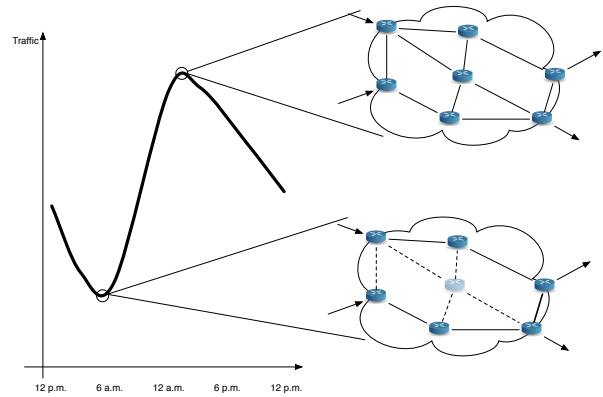

Fig. 5. Energy-aware routing: routers are put to sleep when the network load is low, while preserving connectivity. This technique may increase the load on some links (which are represented with different edges thicknesses on the picture) and QoS performance need to be carefully studied.

*B. Energy-aware routing*

If much improvement is expected from the link layer, through link adaptations and proxying techniques, the network layer may also be involved in the reduction of the energy expenditure. Following the resource-consolidation principle, Energy-aware routing (illustrated in Fig. 5), generally aims at aggregating traffic flows over a subset of the network devices and links, allowing other links and interconnection devices to be switched off. These solutions shall preserve connectivity and QoS, for instance by limiting the maximum utilization over any link, or ensuring a minimum level of path diversity. Flows aggregation may be achieved, for example, through a proper configuration of the routing weights. Formally, energy-aware routing is a particular instance of the general capacitated multi-commodity flow problems [50], and thus falls into the set of the off-line solutions concerning dimensioning.

Energy-aware routing has been first evoked in the position paper [19], but just as a hypothetical working direction (under the name of "coordinated sleeping"), by taking the example of two parallel routers acting on the boundaries of an AS. The possibility of coordinating the sleeping periods of these two routers is discussed, speculating on the impacts on fault tolerance and the required changes to routing protocols. On the one hand, OSPF considers, by default, sleeping links as faulty, and requires an update of the shortest paths, which requires time consuming computation. On the other hand, IBGP may suffer from routes oscillation and see occasionally forwarding loops. The paper indicates a possible solution to this problem through the use of different precomputed solutions, or in the presence of a unique centralized omniscient decision point.

In [50], the problem is formulated using Integer Linear Programming (ILP), and some greedy heuristics are examined, which progressively switch off nodes and links. The authors study the effects of different selection strategies on the set of nodes and links. The problem is handled at the level of ISP networks, with multi-homed edge nodes: as results on more general topologies are not provided, this can be considered a sort of best-case scenario (since having a relevant number of redundant nodes and links considerably increase the efficiency of the solution). Other work such as [51], [52] instead solve the ILP problem numerically, by considering that only links [51] or both links and nodes [52] can be turned off.

More generally, in the context of energy-aware routing an analysis of the effects on the inter-domain traffic is still missing, as well as a study on the robustness of the solution to faults and traffic changes.

## VIII. ENERGY-AWARE SOFTWARE AND APPLICATIONS

Modifying the network operation according to the varying load imposed by the applications represents, as examined above, a good opportunity to avoid unnecessary energy waste. Yet, the very same operating systems and applications can be modified to participate in the reduction of the energy budget.

### A. User-level applications

For instance [39] redesigns the Telnet protocol in a green perspective, allowing the client to go to sleep after a given time and recover later. This requires a modification of the protocol implementation (notification of the clients idle states to the server requires additional signaling), to avoid losing data without sending keep-alive messages.

Green BitTorrent [40] follows the same spirit. Peers openly advertise their energy-state, and green peers tend to avoid waking up idle peers, preferring to download the chunks from active ones. A probing mechanism is defined to test peers whose status is unknown such as the ones advertised by the tracker. However, the correct way to signal peers energy status without requiring an active beaconing system is not mentioned in this publication, nor are methods needed to wake up the clients from their idle state. Authors of [40] suggest to use Wake on LAN, which however poses security issues. Otherwise, green functionalities could be included in the set-top-box, an area that needs to be further explored.

More generally, tools such as the ones proposed by [53], [54], targeting application energy profiling and energy-aware programming, can assist development of green user-level applications.

### B. Kernel-level network stack

Other than greening the application-layer in the user-space, it is also possible to improve the transport-layer at kernel-level for more energy-efficiency. The benefit of this approach is that these optimizations are then shared by all applications. More specifically, modification of TCP in the operating systems kernels, would allow applications to open "greener" sockets, providing a framework for software developers. The author of [37] suggests one such modification, introducing explicit signaling at the transport layer via a specific option (TCP_SLEEP) in the TCP header, in which case the other party will buffer data received from the application instead of sending it right away. This strategy is illustrated in Fig. 6, and an actual implementation involving more details such as the definition of a maximum allowable sleep time is presented and evaluated in [37].

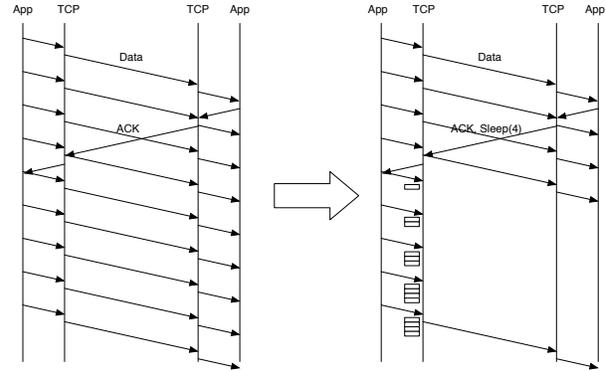

Fig. 6. In a modified version of TCP (right), the receiver may notify its peer of its intention to sleep. During the sleep period (expressed here in number of segments for the sake of illustration), the source buffers the specified amount of data instead of directly transmitting it.

Orthogonally, other contributions examine the energy consumption induced by the computational load of TCP over FreeBSD and Linux system, aiming at enhancing the energy efficiency of existing implementations rather than modifying the protocol stack [38]. The authors present a detailed breakdown of the cost of TCP/IP networking between operation in the NIC, kernel and user space, estimating to about 15% the TCP processing cost, among which about 20% to 30% is due to the single checksum computation. They also propose further solutions to reduce the energy consumption by TCP, by exploiting some functionalities at the NIC level.

## IX. AT THE NETWORK EDGE

Finally, we briefly overview the current research trends in areas that are closely related to, but out of the strict scope of the current survey – namely, *computer* and *data-center* architectures and *wireless* networking.

### A. Computer and Data-Center Architectures

End-user computers and data-centers represent the entities that are interconnected through the network, and we argue that energy-efficiency of the end user and of service provider is of important concern for the overall energy budget – especially due to the large number of devices involved. It should be noted that, due to the difference in these architectures, the network-oriented taxonomy of this paper is not relevant for these works.

Let us focus on computer architectures first. Starting from [55] (which however targeted mainly battery conservation in a wireless network), operating systems have undergone a systematic exploration, touching many aspects of their duty, from disks, to security, including CPU scheduling and memory allocation. Energy-awareness in a single system include techniques such as dynamic voltage scaling [18], [56], techniques to group instructions to increase scheduling logic efficiency [57], energy-aware memory allocation [58], specific thread migration policies [59] or energy management policies [60] to increase energy efficiency in multi-core architectures. Effort to bring all the relevant features into operating systems aware of the network and of the energy resources is undergoing [61]. Some of these techniques are already starting to be deployed as tick-less kernel (dyntick) [62] available as a patch since Linux kernel version 2.6.17, which avoids system periodic wake-ups every few milliseconds, allowing the systems to stay idle for long intervals.

With respect to green networking, efficiency in datacenters is a more confined and mature field, which dates back to early 2000 [63]. As previously outlined, green paradigm of selective connectedness and resource consolidation are applied also in data-center research: e.g., [64] jointly applies ALR with Network Traffic and Server Load Consolidation techniques. It can be expected that more techniques can be exported across these fields in the coming years. For further references and an updated state-of-art we refer the reader to [65], pointing out that most recent evolution of energy efficient datacenter research include the tendency to break down monolithic datacenters into distributed nano-datacenters [66], [67].

*B. Wireless communications*

As opposed to the wired environment, where energy-related issues have only become a subject of investigation recently, energy-awareness has been a primary concern in the design of wireless communications since the advent of mobile devices.

First, most mobile devices have limited processing capabilities and operate using limited peak voltage batteries, which imposes a stringent limitation on the instantaneous energy consumption they can support. Moreover, as soon as devices autonomy becomes a concern, the radio interface behavior shall be optimized, as both radio signal emission and reception have a high energy cost. Therefore, as highlighted in [17], most mobile devices and wireless sensors tend to have a low duty cycle, operating by means of short periods of intense performance, followed by long idle intervals.

In wired networks, some edge devices may follow this kind of duty cycling, but most infrastructure network devices may rather exhibit a low average level of utilization and can only afford very short intervals of complete idleness. Therefore, most of the research on energy efficiency in wireless networks is not easily transposable to wired networks. For a longer discussion on these aspects, we refer the interested reader to dedicated surveys, for instance [68].

In addition, wireless communications are prone to collisions and interferences due to the broadcasting nature of the radio medium. Therefore, raw performance and energy savings objectives meet in this context, as a careful transmission scheduling prevents these phenomena (which represent a waste of channel bandwidth and energy). In wired communications, as links are generally isolated, these two objectives are, on the contrary, in opposition. Therefore, a slight part of the wireless techniques are specific to this particular medium and are not directly transferable to the wired infrastructures.

However, while at first sight the wired and wireless environments have orthogonal characteristics, some adaptations of the techniques used in either context is possible and have already been attempted. The best example is represented by Dynamic Voltage Scaling (DVS) [18]: DVS has originally been motivated by the need to reduce the energy consumption of battery-operated systems in the wireless environment, and is nowadays largely used in many computer architectures outside the wireless environment. Notice that also Adaptive Link Rate (ALR) techniques described in Section V share some similarities across both worlds: similarly to what happens in IEEE 802.3az, also in IEEE 802.11 a lower data rate implies a less complex physical layer modulation, thus requiring less power-consuming decoding processing in the electronics domain.

Other similarities exist as well: concerning cellular networks, there is a limit, due to the scheduling policy, in the number of users that a base station can serve. In dense urban areas, these networks are therefore dimensioned with a dense mesh of low-range base stations, which increases the maximum number of simultaneous parallel communications. At night, this results in a network over-provisioning, similar to what happens in the Energy-aware infrastructure category described in Section VII. For instance, [69] proposes to switch off part of the base stations when the network load decreases and to increase the transmission range of the other base stations to preserve coverage, adopting the same heuristic used in wired network [50]. Thus, while using specificities of the wireless link, this technique clearly fall in the Energy-aware infrastructure category according to our taxonomy.

Finally, besides these specific contributions, Interface proxying techniques are independent of the type of link and can be applied in wireless networks as well. Energy-aware applications behavior may be tuned according to the type of link they operate on (for instance, energy-aware TCP may send more frequent SLEEP requests when using a wireless link), although this may requires cross-layer solutions that break the OSI layers independence assumption.

## X. MEASUREMENT AND MODELS

As outlined in the previous sections, several works propose some modifications to the link layer, networking layer, transport layer and applications themselves, in order to improve energy consumption in the network. As the OSI model supposes independence of the different layers, it seems natural to develop these optimizations in an uncoordinated manner. However, some of these proposals may be more effective than others, and the interaction between the techniques may be counter-productive. There is therefore a real need for a common evaluation framework and methodology.

Some effort has been directed toward power modeling and power measurements. Models and measurements not only motivate green networking principles, or assessing their feasibility, but are indeed also necessary to evaluate the performance of green solutions. The next paragraphs survey the literature concerning the energy consumption evaluation of single and networked systems, before discussing metrics and benchmarks.

### A. Finding new green opportunities

First of all, empirical observations from measurement of the behavior of networks and applications represent a key requirement to assess the feasibility of green networking principles.

The feasibility of proportional computing approaches, for instance, is strongly influenced by the real traffic patterns, as shown by a number of works including [19], [21], [31], where the analysis of packet inter-arrival times into line cards was first used to motivate the case for sleeping during idle periods. The analysis of the amount and characteristics of chatter traffic [21], [31] on the other hand motivated the selective connectedness principle and proxying techniques. Measurement studies capturing network traffic patterns and user presence indicators such as the one conducted in [31] are also necessary to evaluate the potential for energy savings due to proxying techniques. Therefore, measurement will likely remain a core tool to identify new, practical, opportunities where energy saving is possible.

### B. Consumption of end systems

Traditionally, energy measurements have been carried out for single, non-networked, computer systems. Such measurement are either component-wise, focused on specific aspects of a systems such as the CPU [56], the memory [58] or the hard-drives [70]. Some works also consider the computer system as a whole [71], [72].

Even if this area is relatively mature, there are some inconsistencies between energy models. Comparison works such as [71] highlight a clear tradeoff between simplicity and accuracy of the energy models, and push towards finer-grained energy models, taking into account detailed reporting for each component (OS, CPU, etc.). The authors of [72] instead propose to use a small set of tightly correlated parameters (such as the processor frequency and consumption, the bus activity or the system ambient temperatures) in order to create a linear regression model relating system energy input to subsystem energy consumption.

Furthermore, software design choices may also have their impact on energy consumption of hosts, both at the kernel and application levels. For instance [53] proposes a fine-grained measurement tool that can assist green-application design. The authors demonstrate the use of their tool through the evaluation of a program that works on compressed data, versus the same program running on uncompressed data. For this particular software, the use of compressed-data introduces a higher CPU load, but a shorter disk usage, resulting in a higher overall energy efficiency. Thus, the authors argue that similar tools should be deeply embedded into the development process to make energy profiling of the application easier at early stages.

### C. Consumption of network devices

Energy consumption figures of network equipment such as switches and routers are generally not available with the correct level of granularity. The datasheets of these devices often indicate a single energy consumption figure, which corresponds to a particular operational mode, or to a maximum energy consumption.

Concerning more detailed measurements and evaluations, a few specific routers were tested in [34] (namely, a Cisco 7500 and a Cisco GSR12008). Along a similar line, [43] measures the energy consumption of four switches, considering both home and professional equipment type, confirming that almost irrespectively of the traffic handled, network appliance energy consumption can be approximated as a constant.

However, the community still lacks a set of representative measurement figures: further efforts are necessary to gather and publish such a representative database, involving different types of equipment (e.g., DSLAMs, Ethernet Switches, set-top-boxes, etc.) and comparing different technologies (e.g., 1 Gb/s Ethernet vs. 10 Gb/s Ethernet, etc.).

### D. Consumption of networks

The energy consumption of larger scale networks is relatively difficult to measure, as several factors have to be included in this budget (e.g., from device power models, to redundancy of the architecture, to air cooling system, etc.). A few models exist that estimate the energy consumption of the current Internet, though.

In more detail [73], [74], provide a simple yet relevant model, which considers several types of access (e.g., PON, FTTH, xDSL, WiMax, etc.) insisting on an optical core. However, some simplifying assumptions need further discussion. For instance, a constant factor of two is used to derive the consumption of cooling and heat dissipation mechanisms, which does not take into account advances in these areas. Moreover, the redundancy of network equipment is neglected, which is a strong assumption as the redundant devices are not fully passive. Finally, the author by design proposes a model to evaluate a lower bound of energy consumption – thus, a similar model for a tight upper bound consumption may be needed to better refine the picture.

### E. Benchmark and metrics

Even if the green networking area has been quite active lately, there is still an urgent need for reliable, representative and up-to-date figures regarding energy consumption. Several contributions proposed some benchmarking techniques, which should be deployed over state of the art solutions so that results can be exchanged and compared. In the computing field, a benchmark has been proposed [75], taking into account energy consumption for data-center operations, and therefore adding a new dimension in the network architecture design. Yet, [75] is based on the external sort benchmark, and is thus more suited for data-center/grid environment than for networks.

Some methods have also been proposed in the literature to evaluate the energy/utilization footprint of a networked device. These methods are closely related to the representation of the energy consumption introduced in Fig. 2(a). In particular, [44] proposes a method to assess how far is a given device from a proportional footprint, and define a scalar metric called Energy Proportionality Index (EPI). However, as EPI abstracts the slope of the proportional devices, it cannot discriminate for instance single-step solutions and multi-step curves. A complementary method is proposed in [17], which instead takes into account the efficiency of the device with respect to the device utilization. Further metrics are collected and compared in [76].

Moreover, it has to be pointed out that although several works adopt the same evaluation technique, the performance of the different approaches can rarely be compared: consider for instance the case of trace-driven evaluatio, which has been used in [19], [25], [20], [21], [22], [29], [27] and where each approach has been tested using different inputs. Rather, in order for the relative merits of different approaches to be more easily and directly comparable, a *standard set of benchmark workloads* should be identified and agreed by the community. A question that needs to be addressed is how to define standard meaningful workloads for the different branches of work, so that system energy consumption can be weighted on the ground of the offered throughput (instead of carrying on comparison based on the capacity, as otherwise an over-provisioned non-reactive system could be seen as very efficient).

Also, consensus should be reached on a *standard set of metrics* used to express performance as well. For instance, let us consider the case of green switching architectures implementing Adaptive Link Rate, such as [25], [23], [24]. In the above works, green performance is expressed in [25] as the reduction in the total energy, whereas [23] and [24] express performance in terms of the percentage of time spent in a low-energy state. Although both metrics are relevant, the scientific community should agree on common ones, as otherwise the use of diverse metrics would again hinder the possibility of direct comparison across proposals. A preliminary effort in this direction is made in [76], that overviews and compare several metrics used in the green networking and datacenter literature.

## XI. CONCLUSION

This article surveyed the efforts that the research community has been spending in the attempt to reduce the energy waste in fixed networks, which are usually denoted as "green networking". We presented the importance of the issue, its definition and mainstream paradigms, and proposed a taxonomy of the relevant related work.

Examining the state of the art, we observe that a few techniques are emerging, which can be roughly categorized as (i) Adaptive Link Rate, (ii) Interface-Proxying, (iii) Energy-Aware Infrastructure and (iv) Energy-Aware Applications.

It also emerges from our analysis that despite the relative youth of the green networking field, research in some of the above areas is already mature, with advanced standardization efforts and prototyping results. In the following, we summarize the specificity of each branch, commenting on their respective level of maturity and providing future research directions.

Notice that the order of each category roughly follows their level of maturity, with Adaptive Link Rate being the most explored area, and Energy-Aware Applications the least mature. Therefore, it naturally follows that more work remains to be done in the less explored area, on which we provide thus further perspective and open problems.

### A. Adaptive Link Rate

In more detail, the first category regroups all the proposals to reduce energy consumption by automatically adapting the link rate to the traffic level.

This class of solutions is already well advanced: the IEEE standardization committee had already produced a first draft [41] at the time this survey was written, therefore products can be expected soon. In this class of work, different strategies have been analyzed, concerning e.g., the rate switching policy (e.g., when to switch, for how long, what criteria should the decision be based on, etc.), or the preferable rate definition (e.g., *sleeping mode* versus *rate switch*, how many rates are necessary and which ones, etc.).

As this area of research has already been largely explored, future work will likely have to thoroughly contrast the different proposed strategies, as a comprehensive view at both local and global levels is still missing to date. Comparison is indeed an important point, which must furthermore take into account not only the energy efficiency issue, but also the effects on the perceived QoS, which we discuss more deeply later on.

### B. Interface Proxying

The second category, namely Interface Proxying, comprises approaches where an intermediate low-power entity handles a number of network operations, offloading the main processor and thereby extending the duration of idle-periods of power-hungry devices.

From our analysis, it emerges that the intermediate entity could be located directly in the network interface card NIC of the main device (thus serving a single client), or be a stand-alone external entity (possibly serving multiple clients). Work in the literature considers several degrees of complexity of the off-loading operations, ranging from simple filtering of chatter traffic (e.g., ARP, ICMP, broadcast, keep-alive, etc.), to more complex handling of non-interactive traffic (e.g., background FTP or P2P-filesharing transfers). From our analysis, it also emerges that research in this area is in an advanced state as well: indeed, several prototype implementations (both on the NIC of the same system or as external devices) have already started to appear, and standardization effort is also ongoing in ECMA [42].

In this area, comparison work does not exist yet, and a thorough analysis of the side effects on QoS tied to system wake-up time is still missing. Moreover, the community seems to have neglected set-top-boxes so far, which will be the most likely candidate for proxying in residential environments. Other interesting research directions come from coordination of different subsystem to further extend the sleep period (hence increasing the energy saving). Notice that the subsystems needing coordination possibly come from different research areas, or need cross-layer interaction. For example, in multi-homed terminals, it should be avoided that the different cards could wake the system up at different times, as synchronization of wake-up times could yield to prolonged sleep intervals: in its turn, this translate into coordinating proxying activities either locally (e.g., several NIC proxies requiring inter-process communication) or remotely (e.g., several external proxies coordinating via a network protocol). Also, end-system wake-up can happen due to multiple reasons (e.g., kernel-level ticks, due to periodic application activities, due to network interrupts raised by packets reception at the NIC, due to periodic heartbeat activities coming from applications that need to send packets to the NIC, etc.), which may require further layers of local and remote coordination.

*C. Energy-Aware Infrastructure*

A third category regroups solutions related to the design and management of Energy-Aware Infrastructures, adopting thus a global network perspective.

Solutions in this field either advocate a clean-slate redesign of the network architecture, or try to incorporate energy-awareness in the family of currently deployed routing protocols. However, due to the intrinsic complexity of deploying radically new architectures (and even to gradually evolve existing architectures), this area of research is less advanced with respect to the former ones.

As a result, many interesting problems remain open at both architectural and operational levels. For example, an unexplored perspective could consider the infrastructural problem from a multi-layer (e.g., opto/electronic tradeoff) point of view. Another interesting open problem is to quantify the potential benefits in designing new energy-efficient architectures, with respect to exploiting current architectures in an energy-efficient way. As far as the operational level is concerned, we observe that while a number of solution propose to reduce the number of active elements, the focus has generally been on minimizing the network energy expenditure, only minimally taking into account the solution robustness (e.g., with a linear combination of objectives [35], or by bounding the maximum link utilization [33], [52]): thus, a more systematic exploration of the QoS tradeoff is required. Moreover, while much effort has been directed in gathering *optimal* solutions, no work has put the accent on gathering *practical* solutions as well (e.g., consider the case where two different Energy-aware routing solutions yield to very similar energy expenditure: in case the highest gain can only be obtained with a complete network reconfiguration, the solution requiring minimal changes may be preferred). Finally, we observe that no work has so far attempted to precisely define the required changes to the existing routing protocols, in neither the single or multi-domain context, which thus remain a largely unexplored research area.

*D. Energy-Aware Applications*

The last category includes research efforts that incorporate energy-awareness in the software design.

Our analysis shows that approaches to make software less eager in terms of the energy it consumes have been attempted at several level. At the lowest level, we have solutions that will be beneficial to all the software relying on them: specifically, low-level software relates to both general aspects of computing architectures (such as operating system with tickless-kernels) or solutions more closely related to the network domain (e.g., modification to the operating system network stack). Greening of the networking stack has targeted both optimization of the current implementations, and protocol design by explicitly including sleep-states. Similarly happens at a higher level, where changes can either pertain to optimization of energy-efficient instructions, or to protocol modification (e.g., as for the green versions of Telnet and BitTorrent).

Interestingly, despite few work has been carried out in this context, however the literature contains a good coverage of the possible approaches (e.g., kernel vs. user, protocol tweak vs. protocol design, client-server vs. P2P applications, etc.). Still, as the spectrum of Internet applications is clearly a huge one, much of the work remains to be done. On the one hand, we believe that advances in this field can happen only provided that specific tools can assist (and automate) the energy measurement and profiling of networked applications. On the other hand, there is need to translate the green principle into good programming practices: to give a simple example, the consolidation principle may push into launching a set of routine network tasks at the same time (eventually with different priorities), rather than trying to spread them out as possible (which instead implicitly limits the maximum achievable sleep time). Similarly, pacing of network packets could be considered a non-green practice, while the typical burstiness due to TCP batch arrivals could be seen as another incarnation of the same resource consolidation principle. Clearly, the impact of green applications could be massive only once the green wave hits both widespread programming libraries and popular end-user applications.

*E. Measurement and Models*

Finally, as far as energy measurement and modeling are concerned, we observe that green networking has given a novel input to this area: we expect this practice to hold for new, so far undiscovered, branches of the green research domain.

Besides making the case for new green research directions, a first major direction for future work lays in the effort for building an early consensus of benchmarking methodology, workload and metrics. In other words, while common ground on benchmark methodologies, input and metrics is instrumental to promote cross comparison of different work, advances in power-modeling are needed to gather reliable results, representative furthermore of realistic scenarios.

Another interesting, and relatively less explored aspect concerns, e.g., the energy-modeling of the Internet, which is currently under investigation but that has not been fully clarified yet. Indeed, as we already pointed out, gathering precise figures is difficult due to the scale of the network on the one hand, and on the fast paced technological evolution. Yet, understanding where the major energy expenditure happens is crucial to pinpoint where the larger energy saving could be obtained.

*F. QoS considerations*

Finally, as the ultimate goal of networking is to provide services to end-users, the quality of such services and of the user experience is a topic that spans over all the previous branches. Indeed, while energy efficiency is becoming a primary issue, it shall never be neglected that the energy gain must not come at the price of a network performance loss.

This delicate tradeoff arises from opposite principles: indeed, while networked systems have traditionally be designed and dimensioned according to principles such as *over-provisioning* and *redundancy*, green networking approaches praise opposite practices such as *resource-consolidation* and *selective-connectedness*. The challenge lays in this case in applying the latter principles in a way that is as transparent as possible to the user – in other words, avoiding that resource-consolidation translates into congestion, or that selective-connectedness translates into unreachability.

While the first wave of green studies focused more on the achievable energy gain, we believe that a systematic evaluation of networking performance from the user-perspective should be undertaken as well. Indeed, in all branches interesting questions remain, which deserve *precise quantitative* answers: e.g., how does the different Adaptive Link Rate techniques compare in terms of additional delay and jitter? What about security issues in Interface Proxying, path protection and resilience in Energy Aware Routing, and the increase of the completion time of green BitTorrent or other Energy Aware Applications?

Finally, we believe that while, for the time being, techniques of different branches have been studied in isolation, future research should address the *combined impact* of different techniques as well. Indeed, even though each of the above techniques alone do not constitute serious threats for the QoS perceived by the end-user, however it is not guaranteed that the joint use of several technique will not raise unexpected behavior. Due to the current rise in green networking research and attention, it cannot be excluded that, in a near future, users will run Energy Aware Applications, in a home equipped with a green set-top-box implementing Interface Proxying functionalities, and will access the Internet through an Internet Service Provider implementing Energy Aware Routing in devices interconnected by Adaptive Link Rate lines – which opens a number of interesting questions that are so far all unexplored.

ACKNOWLEDGMENT

This work was funded by TIGER2, a project of the Eureka Celtic Cluster Framework program.